\documentclass[a4paper,UKenglish,cleveref, autoref, nolinenumbers, thm-restate]{lipics-v2021}
\usepackage{amsmath}
\usepackage{amssymb}

\usepackage{microtype}                 % use micro-typography (slightly more compact, better to read)
\PassOptionsToPackage{warn}{textcomp}  % to address font issues with \textrightarrow
\usepackage{textcomp}                  % use better special symbols
\usepackage{mathptmx}[osf]                  % use matching math font
\usepackage{times}                     % we use Times as the main font
\usepackage{cite}                      % needed to automatically sort the references
\usepackage{tabu}                      % only used for the table example
\usepackage{booktabs}                  % only used for the table example
\usepackage{hyperref}
\usepackage{cleveref}

\usepackage{balance}
\usepackage{graphicx}
\usepackage{framed}

\usepackage{color}
\usepackage{subcaption}
\usepackage{wrapfig}
\usepackage{enumitem}
\usepackage{xspace}
\usepackage{xcolor}
\usepackage{colortbl}
\usepackage{mathtools}

\definecolor{blue}{HTML}{0066cc}

\newcommand{\red}[1]{\textcolor{red}{#1}}

\newcommand{\blue}[1]{\textcolor{blue}{#1}}

\newlength{\listingindent}                %declare a new length
\usepackage[LGRgreek]{mathastext}

\DeclareFontFamily{U}{matha}{\hyphenchar\font45}
\DeclareFontShape{U}{matha}{m}{n}{
      <5> <6> <7> <8> <9> <10> gen * matha
      <10.95> matha10 <12> <14.4> <17.28> <20.74> <24.88> matha12
      }{}
\DeclareSymbolFont{matha}{U}{matha}{m}{n}

\DeclareMathSymbol{\Lt}{3}{matha}{"CE}
\DeclareMathSymbol{\Gt}{3}{matha}{"CF}

\setlist{leftmargin=*,itemsep=0pt}

\DeclareMathAlphabet{\mathit}{T1}{cmr}{m}{it}

\newcommand{\stitle}[1]{\smallskip\noindent\textbf{#1}}

\usepackage{lineno}

\title{Database Theory + X: Database Visualization} %TODO Please add

\author{Eugene Wu}{Columbia University, USA}{ewu@cs.columbia.edu}{https://orcid.org/0000-0003-4254-6688}{} %fundings }
\authorrunning{E. Wu} %TODO mandatory. First: Use abbreviated first/middle names. Second (only in severe cases): Use first author plus 'et al.'

\Copyright{Eugene Wu Open Access} %TODO mandatory, please use full first names. LIPIcs license is "CC-BY";  http://creativecommons.org/licenses/by/3.0/

\ccsdesc[500]{Human-centered computing~Visualization theory, concepts and paradigms}
\ccsdesc[500]{Information systems~Database design and models}
\keywords{Visualization Theory, Data Model, Database Visualization} %TODO mandatory; please add comma-separated list of keywords

\category{} %optional, e.g. invited paper

\relatedversion{} %optional, e.g. full version hosted on arXiv, HAL, or other respository/website
\funding{This material is based upon work supported by NSF 1845638, 1740305, 2008295, 2106197, 2103794, 2312991, Amazon, Google, Adobe, and CAIT.  Any opinions, findings, and conclusions or recommendations expressed in this material are those of the author(s) and do not necessarily reflect the views of the funders.}%optional, to capture a funding statement, which applies to all authors. Please enter author specific funding statements as fifth argument of the \author macro.

\acknowledgements{Thanks to Wolfgang Batterbauer and Eirik Bakke for early feedback on this work. }%optional

\nolinenumbers %uncomment to disable line numbering

\EventEditors{John Q. Open and Joan R. Access}
\EventNoEds{2}
\EventLongTitle{42nd Conference on Very Important Topics (CVIT 2016)}
\EventShortTitle{CVIT 2016}
\EventAcronym{CVIT}
\EventYear{2016}
\EventDate{December 24--27, 2016}
\EventLocation{Little Whinging, United Kingdom}
\EventLogo{}
\SeriesVolume{42}
\ArticleNo{23}
\begin{document}

\maketitle

%TODO mandatory: add short abstract of the document
\begin{abstract}
We draw a connection between data modeling and visualization, namely that a visualization specification defines a mapping from database constraints to visual representations of those constraints. 
%Using this formalism, 
We show data modeling explains many existing visualization design and introduce multi-table {\it database} visualization. 

\end{abstract}

%% the only exception to this rule is the \firstsection command
\section{Introduction}
\label{s:intro}

Existing visualization theory, libraries, and graphical grammars assume that a single input table is processed to produce an output data visualization.   Yet, database theory shows that one table does not model all data (without introducing redundancy).
We go beyond the single-table data model towards a theory of multi-table {\it Database Visualization}.  %We focus on relational databases because they are nearly ubiquitous, grounded in decades of theory, and can model common data like graphs and hierarchies.  
The theory models a visualization as a visual representation of database contents (the rows and values) {\it that preserves database constraints}.  We present the main ideas through a series of examples and show how under this theory, some common visualization designs are a consequence of data modeling choices.

%Building on this theory, we use a series of examples to illustrate the connection between data modeling and visualization designs.   We end with potential research questions that may be of interest to the database theory community.  

\if{0}
Relational data theory is based on two related concepts.  The first is the model of a database as one or more tables, and a set of constraints over these table.   These constraints specify valid values for a given attribute (e.g., data types, NOT NULL), as well as relational constraints between attributes in the same table (e.g., zip determines state) and across multiple tables (e.g., xact.custid references customer.id).

This results in many problems that break the compositional abstraction, potentially lead to developer confusion, and make automated analysis of the visualization specification challenging.
Lack of composition
Confusing effects of composition
Non-orthogonal components: facets.
Expressiveness limitations: facets, subplots, grouped bars, node-link diagrams, tree maps and space filling visualizations, working around the encoding abstraction to express parallel coordinates.

%This paper draws the connection

This short paper outlines the connection between data modeling and data visualization, 

However, a key assumption in preceding theory and grammars is that the input is a single table that is mapping to visual marks.     This results in many problems that break the compositional abstraction, potentially lead to developer confusion, and make automated analysis of the visualization specification challenging.
Lack of composition
Confusing effects of composition
Non-orthogonal components: facets.
Expressiveness limitations: facets, subplots, grouped bars, node-link diagrams, tree maps and space filling visualizations, working around the encoding abstraction to express parallel coordinates.

The second is the theory of data modeling, which is the choice of how to decompose one table into multiple tables in order to reduce redundancies as defined by the constraints.   This leads to many benefits---it  avoids potential inconsistencies, improves query evaluation, reduces the amount of redundant data that must be stored, and explains the structure of the data at the schema level---that have correspondences in the visualization domain.

This makes supporting many classes of visualization challenging, and requires embedding complex logic as custom functions or operators within the graphical grammar.    

This paper explores the perspective that visualization is the mapping of both data values {\it and constraints over the data} to a visual representation.
By extending to constraints, .
We show this in the context of constraints under the relational model.

Extending to constraints requires addressing two questions.
First, is what visual representations can constraints be mapped to?
In formalism based on the grammar of graphics, the visual mapping corresponds to mapping data attributes to mark properties (or visual channels).

the correspondence between data modelling---particularly, the concept of normalization and constraints---and its visualization representation, and suggests that a large class of visualizations can be defined based on the underlying data modeling decisions and specifying how the resulting tables and relational constraints map to visual objects.  This describes a broad set of existing visualization and visual-coding designs, as well as live programming and other environments.  It also delineates the role of layout from visualization structure.

We also present an implementation of this formalism.

What is a correct and consistent visualization?
We show that recent proposals for visualizatino consistency are implicitly designed for a degenerate form of consistence between exactly the same attributes.

This formalism is general thanks to the expressiveness of the relational model, which has been shown to faithfully model a wide range of practical data, including tabular, hierarchical, network, and temporal data.

no distinction between mark, subplot, view, and multi-view dashboard.
visualization is a process of mapping constraints and data.
Mechanisms can then be reused.

from scales to layouts: train(data, constraints) then apply.
* decouple data trained on and data applied to
\fi

\section{Background}

This section reviews visualization theory, the relational data model, and single table grammars.  
\subsection{Visualization Theory and the Single-Table Assumption}

\begin{wrapfigure}{r}{.4\textwidth}
    \centering
    \vspace{-.2in}
    \includegraphics[width=.39\textwidth]{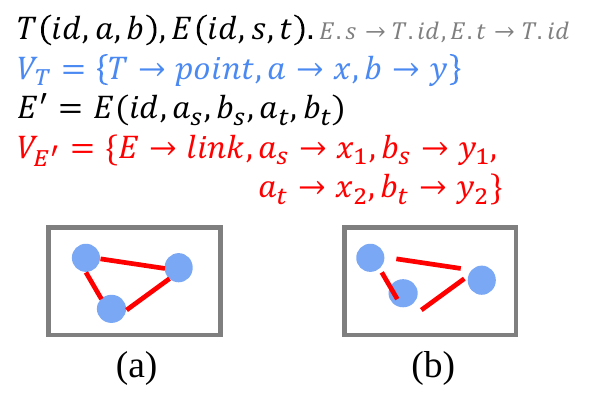}
    \vspace{-.15in}
    \caption{In the node link visualization, (a) the links \red{$V_{E'}$} only appear to connect the points \blue{$V_T$}. (b) The points and links become inconsistent if \blue{$V_T$} changes (e.g., jitter).}
\label{fig:nodelink}
    
\end{wrapfigure}

%The core theory that underpins modern data visualization is based on 
Bertin first described the data-to-visual (or pixel) mapping, where tuples map to marks (e.g., points), and data attributes map to the mark's visual channels~\cite{Bertin1983TheSO}. 
For instance, \Cref{fig:nodelink}(a) visualizes a table $T(id, a, b)$ as a scatter plot \blue{$V_T$} where $T.a$ and $T.b$ are mapped to the x and y axes (ignore the red edges).
Wilkinson's Grammar of Graphics~\cite{wilkinson2012grammar} extended this mapping with scales, mark types, facets, and more to develop a ``graphical grammar'' that defines a design space of data visualizations, which forms the foundation of most data visualization systems including nViZn~\cite{Wilkinson2001nViZnA}, ggplot2~\cite{Wickham2010ALG}, Tableau~\cite{stolte2002polaris}, and vega-lite~\cite{Satyanarayan2018VegaLiteAG}.

However, this theory is predicated on a single-table data model, which does not model most data such as hierarchical JSON, RDF graphs, and relational databases.  Thus, the user is expected to ''prepare'' their data (extract, transform, filter, join) into a single table before visualization.  This is a major burden, cannot guarantee that the visualization faithfully reflects the data, and makes it impossible for existing formalisms to express seemingly common designs such as node-link diagrams, parallel coordinates, tree diagrams, nested visualizations, nor tree maps.  

For instance, \Cref{fig:nodelink} connects the points in  \blue{$V_T$} based on the edges table $E(id,s,t)$.  In existing approaches, the user first computes $E'=T\Join E\Join T$, then maps $E'$ as links in \red{$V_{E'}$}.  However, the edges only {\it appear} to connect the points.  If the user jitters the points or applies a force-directed layout algorithm, then the positions of the points and edges become inconsistent because the visualization is unable to maintain their logical relationship (\Cref{fig:nodelink}(b)).
Furthermore, single-table formalisms and libraries cannot support the force-directed layout algorithm because it needs to reason about {\it two} tables.  As a consequence, the user needs to do more ``preparation'': run the layout to compute pixel coordinates of the nodes in a new table $T'$, derive a new edges table $E''=T'\Join E\Join T'$, and draw the points and links.  This breaks the data-to-pixel mapping abstraction because $T'$ and $E''$ are tables of pixel coordinates, and the grammar is merely used for rendering because all of the layout logic was applied during preparation.      

%\ewu{* You talk about the relational data model as another thing to try to visualize. I.e. the input is now multiple tables of data rather than a single table of data. You could pick this as a starting point without necessarily justifying it, or you could compare with e.g. JSON and RDF and give an argument for why the relational data model is a good starting point. To go from the relational data model to JSON you need an entire query language, such as GraphQL. (SIEUFERD/Ultorg's visual query language is essentially also a relational-to-JSON converter.) You'd need a JSON-like model to visualize e.g. CRUD forms, but you don't need it for e.g. the node link diagrams you mention. }

%The intuition for this limitation draws from data modeling: most applications (including many visual analyses) simply cannot be expressed using a single table.  Instead, the majority of modern databases follow the relational model~\cite{codd1970relational} which informally stipulates that the database consists of multiple tables {\it as well as} the relationships between those tables.   

\subsection{Relational Data Modeling}

We denote a table by its schema $T(a_1,\dots,a_n)$, where attribute $T.a_i$ has domain  $T.\mathbb{D}_i$, and use capitalization to denote a set of attributes $T.X$.   We underline a key $\underline{T.X}$, and assume every table has primary key $\underline{T.id}$.
A foreign key relationship $C(S.X,T.Y)$ specifies that a subset of attributes $S.X\subseteq S_S$ refer to attributes $T.Y\subseteq S_T$ of another table\footnote{Codd's definition~\cite{codd1970relational} is more strict, and  states that $S.X$ is not a key and $T.Y$ is a primary key.  We relax this definition in order to model relationships as used in practice.  }:  $\forall s\in S \exists t\in T,\ s.X = t.Y$.  
1-1 relationships are when both $S.X$ and $T.Y$ are keys, denoted $C(\underline{S.X},\underline{T.Y})$; N-1 relationships denoted $C(S.X,\underline{T.Y})$.  N-M relationships are modeled using an intermediate relation $W(X,Y)$ with constraints $C(W.X,\underline{S.X})$ and $C(W.Y,\underline{T.Y})$.

\subsection{Single-Table Grammars}
We now introduce a simple formalism to describe graphical grammars that visually map an input table $T$ to an output view $V$ (a {\it visual mapping}). 
The notation focuses on data attributes, marks, mappings between data and mark properties, and scales.  
A visualization may have multiple views, often laid out next to each other, or super-imposed as layers.

In database parlance, the view $V$ is a projection of data attributes to mark properties (often called a visual channel) that is indexed by the table's keys.     
%A view $V$ maps input table $T$ to a mark type, each row in $T$ to a mark indexed by the table's keys, and data attributes to the mark's properties (often called a visual channel).    
The notation $a\overset{s}{\to}v$ specifies an {\it aesthetic mapping} from data attribute $a$ to mark property $v$ using a scale function $s$.    Named scale functions can be referenced in multiple views; by default each mapping uses a different scale function.
For instance, the following creates a scatterplot that uses linear scales to map values of $a$ to pixel positions from $10$ to $100$ along the x axis (similarly for $b\overset{s_y}{\to}y$, and the primary key $id$ using an identity function ($s_{id}$).  
\begin{align*}
     V = \{ T\to point,\ T.id\overset{s_{id}}{\to}id,\ T.a\overset{s_x}{\to} x,\ T.b\overset{s_y}{\to}y \}\hspace{3em}
     s_{id} = identity(), s_x = linear(),  s_y = linear() 
    %s_{id} &= identity\hspace{3em}
    %s_x = linear\hspace{3em}
    %s_y = linear
\end{align*}
$V$ is a table of marks with schema $V(type,\underline{id},x,y,\dots)$ that contains its mark type, keys, and mark properties.   $V$ has a 1-1 relationship with $T$ via the constraint $C(\underline{V.id},\underline{T.id})$, so it preserves the same relationships that $T$ has with other tables. 
We will use a simplified notation that omits the primary key mapping, table prefixes, and scale functions, unless they are important to an example.  For instance, the following is equivalent to the above example: $V = \{ T\to point, a\to x, b\to y \}$.

\section{Database Visualization}

{\it Database visualization} maps database contents {\it and constraints} to a visual representation.   We describe single-table grammars as  constraint-preserving mapping, introduce extensions to foreign-key constraints, and illustrate the connection between data modeling and visualization design. 

\subsection{Graphical Grammars as Mapping Constraints}\label{ss:mapkeys}
We first examine the existing single-table formalism from the perspective of preserving attribute domain and key constraints.  Attribute domains are preserved by drawing axes or legends to illustrate how the data attribute's domain maps to the range of the visual channel.     Without these visual metadata, marks are simply objects floating in space, and the user cannot interpret the visually encoded data.% the visualization.

To understand key constraints, let us revisit the scatterplot in the previous section: 
$V = \{T\to point, id\to id, year\to x, price\to y\}$.
Te preserve the key constraint $\underline{T.id}$, we must be able to distinguish different rows in the visualization.  
Is the constraint trivially satisfied because $\underline{V.id}=\underline{T.id}$?  Unfortunately no, because the visualization is not a set or bag of rows in an N-dimensional space---its marks are spatially positioned in a 2D space.  Thus, different marks may appear indistinguishable and thus violate the constraint.    This is commonly called ``overplotting''.
%
    %$$indistinguishable(m_1,m_2)\implies m_1=m_2\hspace{3em}m_1,m_2\in V$$
%

What are ways to address this constraint violation?  
One common approach is to change the constraints by e.g., grouping by the spatial attributes ($T.(year,price)$) to render a heatmap where the spatial attributes are the key, or reducing opacity achieves a similar goal but groups at the pixel granularity.  
A second approach is to change the specification by e.g., mapping $id\to x$ and $year\to color$ to remove overlaps.   The third is to perturb the layout by e.g., jittering the points to reduce overlaps.  
This variety of interventions is possible because the violations are visual and low dimensional. 
 This contrasts with data cleaning, which resolves violations by changing data values.

%\begin{table}
%    \centering
%    \begin{tabular}{rl}
%        \textbf{Data Constraint} & \textbf{Visual Representation} \\
%        Key & \red{Mark per row} \\
%        Attribute Domain& \blue{Visual channel, axis/guide} \\
%        1-1 Relationship& \red{Mark}, \blue{share scale, align, merge marks}  \\
%        1-N Relationship& \red{Mark}, \blue{nest} \\
%        N-M Relationship& \red{Mark} \\
%    \end{tabular}
%    \caption{Summary of visual representations for different database constraints.  \red{Red text} and \blue{blue text} denote explicit and implict representations, respectively.}
%    \label{tab:my_label}
%\end{table}

\subsection{Data Modeling and Visualization Through Examples}

\begin{figure*}
    \centering
    \includegraphics[width=.9\textwidth]{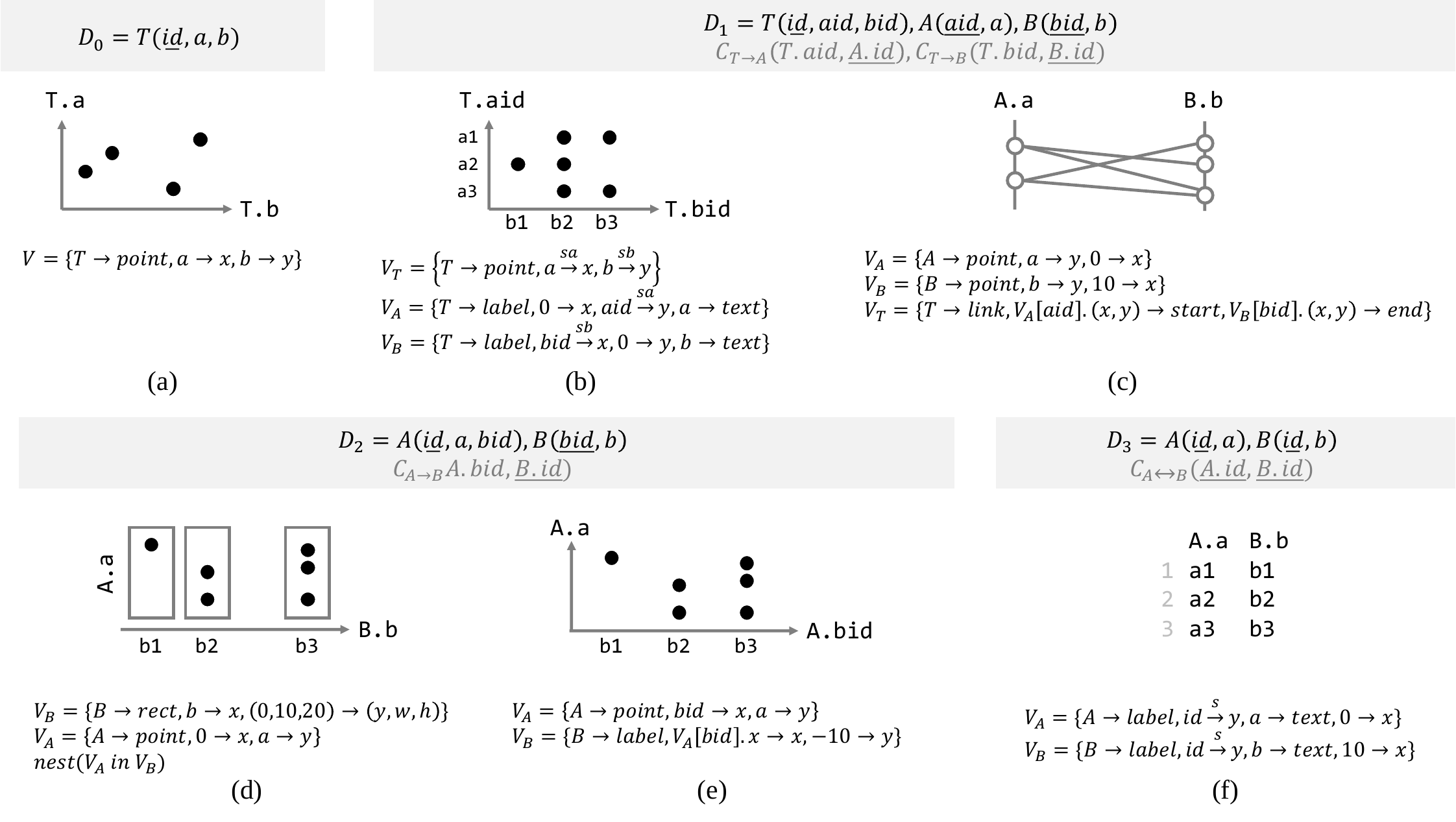}
    \caption{Examples that show how different visualization structures are needed to faithfully express different sets of tables and constraints. Common visualization designs are a consequence of data modeling decisions. }
    \label{fig:gallery}
\end{figure*}

We define a {\it faithful database visualization} as a mapping where 1) each table maps to one or more views; in a given view, 2) each row maps to one mark and 3) each attribute maps to a mark property; and 4) each constraint is preserved in the views.  Single-table mappings are sufficient to express (1-3), so we now shift focus to (4) using the series of examples in \Cref{fig:gallery}.  Starting with the database $D_0$ in \Cref{fig:gallery}(a), we different decompositions to illustrate how the data model affects the structure of the visualization, and several ways to visually preserve foreign keys.

\stitle{Many-Many Decomposition.}
$D_1$ normalizes the attributes $a$ and $b$ as their own entities, and the original $T$ encodes a many-to-many relationship between tables $A$ and $B$, as is common for graph data. 
\Cref{fig:gallery}(b) renders the cross product $T.aid\times T.bid$ as the scatterplot $V_T$ and marks each row in $T$ with a point. $A$ and $B$ are rendered as labels along the y (in $V_A$) and x-axes (in $V_B$); the relationships are preserved by sharing the scales so that the labels are aligned with their positions in $V_T$. For instance $V_A$ maps $aid\to y$ using the same scale $sa$ as in $V_T$ to preserve their correspondence.

\Cref{fig:gallery}(c) visualizes the values of $A.a$ and $B.b$ along separate axes, and renders $T$ as links.  $V_T$ does not directly map $T.aid$ as in \Cref{fig:gallery}(b). It instead uses $T.aid$ to look up its corresponding point mark via $V_A[aid]$; this follows the foreign key constraint $C_{T\to A}$ and the $1-1$ relationship from $A$ to $V_A$, and then maps the point mark's position (which is a key) to the link's start position.  Referencing $V_A[aid]$ via the foreign keys allows them to be maintained in the visualization even as marks move. We note that this design, commonly called parallel coordinates, is a consequence of data modeling.

In general, the specification of a view of $T$ may refer to any table $T'$ for which there is an unambiguous join path $T\leadsto T'$, so that each tuple in $T$ derives a unique value.
In both of these examples, the foreign key relationships are preserved through alignment of scales or mark position.

\stitle{Many-One Decomposition.}
$D_2$ only normalizes $b$, and is commonly used to express hierarchical data.   \Cref{fig:gallery}(d) visualizes $A$ as a scatter plot $V_A$ where the y position is determined by $A.a$, while $B$ is visualized in $V_B$ as rectangle marks whose x-positions are based on $B.b$.   The visualization so far is invalid because $A.bid$ is not mapped and the constraint $C(A.bid,\underline{B.bid})$ is not preserved.   Since each $B$ row maps to multiple $A$ rows, $nest(V_A\ in\ V_B)$ preserves the constraint by drawing $V_A$ within marks of $V_B$.  Specifically, it treats each $V_B$ mark as a ``subcanvas'' and renders $V_A$ inside of it using its related rows in $A$.  Depending on the individual view definitions, nesting expresses faceting, framed-rectangle plots, treemaps, and other nested visualizations.

\Cref{fig:gallery}(e) renders $A$ as a scatterplot where $A.bid$ is mapped to the x position, and $V_B$ renders the corresponding labels for $B$.  This effectively treats $b$ values as a categorical variable, 

%maps the value of $A.a$ to a point's position, similarly to the parallel coordinates visualization, but maps $B.b$ as labels whose positions are equi-spaced based on $B.bid$.   Here, the constraint $C$ is treated as a table and mapped to links that connect related $V_A$ and $V_B$ marks.  The light gray labels show the value of $B.bid$ for the reader's reference.  

\stitle{One-One Decomposition.}
The final decomposition $D_3$ treats $a$ and $b$ as separate entities with a 1-1 relationship.   \Cref{fig:gallery}(f) maps $A.a$ (and $B.b$) to labels where their $id$ determines their positions (similar to $B.b$ in the previous visualization).   Since $V_A$ and $V_B$ are aligned along the $y$ position (derived from their input tables' key), this visualization is faithful.  This is commonly called a table.

\section{Conclusion and Future Work}

This short paper has outlined the basic principles for database visualization as a mapping from database constraints and contents to a faithful visual representation. 
%We started by casting existing visualization formalisms as a mapping from the constraints and contents of a single table to the visualization space, and then extending to foreign-key relationships in order to support relational databases.   
%Through examples, we showed the connections between visualization structure and normalization choices.  
Ultimately, database visualization defines a new design space that encapsulates many visualization designs, such as parallel coordinates and facets, that are not expressible under a single-table formalism.  
This work also raises questions of potential interest for database theorists and practitioners.  
To what extent do the lessons in data modeling---identifying salient entities and relationships, normal forms, cost estimation---translate to and explain visualization designs?
How can layout algorithms (e.g. force-directed layouts) be incorporated into this formalism, and what is the line between declarative mapping specifications and imperative layout algorithms---to what extent are visualizations ``just queries'', really?
Is there a formalism that expresses the continuum between diagrams of the database metadata (e.g., ER diagram) and visualizations of the database contents---does this require a formalism over second-order queries?
A common source of multi-table databases is the intermediate results generated by relational or data science pipelines. What constraints are present in the resulting database of intermediate results beyond those inferred by the chase~\cite{chase}?
What are the visual representations for constraints in non-relational data models such as event streams, matrices, and 3D representations?

%%% if specified like this the section will be committed in review mode
%\acknowledgments{ The authors thank Wolfgang Gatterbauer and Eirik Bakke for insightful suggestions on an early draft. This work is supported by the National Science Foundation under Grant No. 1845638, 1740305, 2008295, 2106197, 2103794, and support from Amazon, Google, Adobe, and CAIT. Any opinions, findings, and conclusions or recommendations expressed in this material are those of the authors and do not reflect the views of the funders.}

%\bibliographystyle{abbrv}
\bibliographystyle{abbrv-doi}

\bibliography{template}

\end{document}